\documentstyle[psfig,floats,epsfig,aps,multicol,fancyheadings]{revtex} 

\def\tc{T$_{c}$}

\def\sb#1{$_{#1}$}

\begin{document}

\draft

\wideabs{

\title{Structural compliance, misfit strain and stripe 
nanostructures in cuprate superconductors}

\author{S. J. L. Billinge and P. M. Duxbury}
\address{Department of Physics and Astronomy and Center for
Fundamental Materials Research,\\ Michigan State University, 
East Lansing, MI 48824-1116.}

\date{\today}

\maketitle

\begin{abstract}
Structural compliance is the ability of a crystal structure
to accommodate variations in local atomic bond-lengths {\it without
incurring large strain energies}.  We show that the structural
compliance of cuprates is relatively small, so that short, highly doped, 
Cu-O-Cu bonds in stripes are subject to a {\it tensile misfit strain}.
We develop a model to describe the effect of misfit strain on 
charge ordering in the copper oxygen planes of oxide materials  
and illustrate some of the low energy stripe 
nanostructures that can result.
\end{abstract}

}

The existence of charge stripes in high T\sb{c}
superconductors~\cite{tranq;n95} and their relevance to the
high-T\sb{c} property~\cite{emery;pnas99,marti;epl01} are hotly debated subjects
at present.  The tendency towards charge phase separation and stripe
formation has been seen in models of strongly correlated electron
systems~\cite{emery;pnas99,marti;epl01,zaane;prb89,white;prb00,stojk;prb00}.  Charge stripes
have also been seen directly in strongly correlated systems such as
manganites~\cite{uehar;n99} and nickelates~\cite{tranq;prl94}.  In the
cuprates, convincing direct evidence for stripes comes with the
observation of charge-order superlattice peaks in
La\sb{1.6-x}Nd\sb{0.4}Sr\sb{x}CuO\sb{4}, though the observation of
static long-range ordered stripes correlates with poor or nonexistent
superconductivity~\cite{tranq;n95,tranq;prb96}.  The interest has
shifted to dynamic fluctuating short-range ordered stripes.  Indirect
experimental evidence exists from inelastic neutron scattering
supporting their presence in
La\sb{2-x}(Sr,Ba)\sb{x}CuO\sb{4}~\cite{kastn;rmp98} and
YBa\sb{2}Cu\sb{3}O\sb{6+\delta}~\cite{mook;n99,arai;prl99}. The
observation of a pseudo gap in the electronic density of states above
T\sb{c} is even more widespread~\cite{timus;rpp99} and can be
interpreted as originating from dynamic stripes.  Of critical
importance is an understanding of what prevents stripes ordering over
long-range and becoming static and insulating.  Here we stress the
importance of lattice strain to this phenomenon.


Strong coupling of doped charges to the lattice is clearly evident in
the cuprates~\cite{tranq;n95,mook;n99,mcque;prl99,lanza;n01}.
A direct way in which the charge couples to the lattice is through the
Cu-O bond length which depends on charge filling of the Cu-O
$\sigma^*$ covalent bond~\cite{goode;f92}.  Doping holes into the
CuO\sb{2} plane reduces charge density in this bond which stabilizes
and shortens the bond.  This is seen experimentally as a shortening of
the Cu-O bond with increasing doping~\cite{radae;prb94i} and has
profound implications when charge-stripes are present.  The presence
of charge stripes implies that regions of the CuO\sb{2} plane that are
heavily doped, and therefore have short bonds, coexist with undoped
regions with long bonds.  Experimental evidence for this coexistence
has been presented in superconducting La\sb{2-x}(Sr,Ba)\sb{x}CuO\sb{4}
samples\cite{bozin;prl00}.  We show that this results in
a misfit strain that breaks up the stripes into a domain
microstructure.  The characteristic length-scale of the domains
depends on the misfit strain and can range from nanometer to long range.
The importance of elastic compliance and interface strain to 
texture formation has been discussed in related ferroelastic 
materials~\cite{sheno;prb99} but this is the first time it
is discussed in the context of breaking up stripes and forming
stripe nanostructures.

A key property of perovskite structures
is their capacity to accommodate atoms of various sizes.
They do this via a variety of buckling and tilt modes.  These modes
have relatively small strain energy costs, and are frequently called
soft or floppy modes~\cite{goode;f92,thorp;jncs83,dove;b;lsfd98}.  In
manganites, nickelates and cuprates, the key soft mode is a buckle in
the Cu-O-Cu bond and doping typically relaxes this buckle in these
materials. This low energy accommodation affects the stability of the
structure against polaron formation~\cite{egami;js00}. We argue that
in cuprates, the short Cu-O-Cu bond is not fully accommodated by
relaxation of the buckle present in the undoped copper-oxygen bond.
This leads to tensile misfit strains in the copper oxygen planes of
cuprates.  In the presence of stripes the tensile misfit strains can
lead to a variety of domain microstructures. We develop a model to
describe these microstructures and present some illustrative results
using the model.  

\begin{figure}
\centerline{\epsfig{file=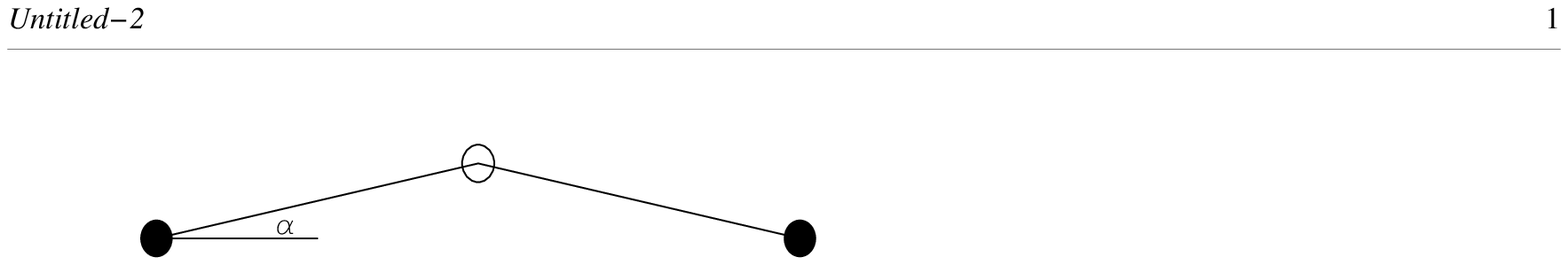,width=6.0cm,clip=TRUE,viewport=90 670 320 720,angle=0}}
\vspace{0.2cm}
\caption{The angle $\alpha$ which describes the buckle in the
copper (solid circle) - oxygen (open circle) - copper bonding.}
\label{fig1}
\end{figure}

The ability of the structure to accommodate local bond shortening we
call ``structural compliance" $\delta_s$.  Where this accommodation
comes from octahedral tilting it is defined as $\delta_s
= {(r_b-r_f)\over r_b}=(1-\cos\alpha)$ where $r_b$ and $r_f$ are the
lengths of the Cu-O bond in the ``buckled" and ``flat'' configurations
respectively and $\alpha$ is the angle the Cu-O bond makes with the
plane (Fig.~\ref{fig1}).
%
%
  In the reduced I4/mmm unit cell of the
  La\sb{2-x}(Sr,Ba)\sb{x}CuO\sb{4} system $r_f$ is simply $a/2$ where
  $a$ is the lattice parameter.  In general $\alpha$ is small and
  $\delta_s \approx {\alpha^2/2}$. The structural compliance therefore
  depends very sensitively on the buckling angle. In the case of
  low-temperature orthorhombic (LTO)~\cite{radae;prb94i}, and
  tetragonal (LTT)~\cite{axe;prl89}, octahedral tilts observed in
  La\sb{2-x}(Sr,Ba)\sb{x}CuO\sb{4},
  $\alpha=\theta_{LTT}=\theta_{LTO}/\sqrt{2}$ where $\theta_{LTT}$ and
  $\theta_{LTO}$ are the octahedral tilt angles in the LTT and LTO
  phases respectively.  We note that for the same tilt angle the LTT
  symmetry tilts have twice the structural compliance.

However the  La\sb{2-x}(Sr,Ba)\sb{x}CuO\sb{4} system has a relatively
small structural compliance, as do the other high T\sb{c} cuprates.
On going from undoped to heavily doped ($x=0.20$)  in the 
 La\sb{2-x}(Sr,Ba)\sb{x}CuO\sb{4} system at 10K, 
the Cu-O bond shortens 
from $r_b=1.904$~\AA\ to $r_b=1.882$~\AA , i.e., by
$\sim 0.022$~\AA~\cite{radae;prb94i}. For illustration purposes, 
we take this bond length
difference to be an estimate of the difference in length
of the short and long Cu-O bonds in the Cu-O planes.
In order for structural compliance to accommodate
 this bond length mismatch we need $\delta_s = 0.022/1.89 =0.012$,
which requires a tilt angle of  8.9$^\circ$ in the LTT symmetry and of 
12.6$^\circ$ in the LTO system.  However
in the La\sb{2-x}(Sr,Ba)\sb{x}CuO\sb{4} system 
tilts have LTO symmetry with a tilt angle of $\sim 3.5^{\circ}$
resulting in a structural compliance of only $\delta_s\sim 0.001$.  There is
therefore insufficient structural compliance to accommodate the
short Cu-O bonds in these materials, resulting in a misfit strain
if charges localize, for example, in stripes.
Note that these numbers represent rough
estimates and the precise value for the crossover will depend on 
the charge density and degree of delocalization of the charge in the
stripe as well as details of whether the stripes are site or bond centered.
Nevertheless, it is evident that in cuprates, stripe formation will lead
to misfit strain, but that materials with moderate tilts 
could accommodate short bonds with little or no misfit strain.


We now consider what happens to the microstructure of the charge
stripes in the case where a misfit strain is present.  We expect/show
that the presence of strain results in a break-up of the stripes into
short segments. Lattice strain therefore prevents the stripes from
ordering over long range; a situation that is a prerequisite for
most stripe theories of high temperature
superconductivity~\cite{emery;pnas99,marti;epl01,white;prl98,%
castr;prb01,caste;zpb97,bussm;jpcm01,bianc;jpcm00}. This has implications for superconductivity
by producing an electronic
microstructure~\cite{phill;pmb99} of domains of broken stripes that results in a high
density of topological stripe defects~\cite{castr;prb01,emery;prl00,zaane;cm01}.
 The ends of the stripes are in a state of high
stress making them especially susceptible to charge fluctuations. 

\begin{figure}
\centerline{\epsfig{file=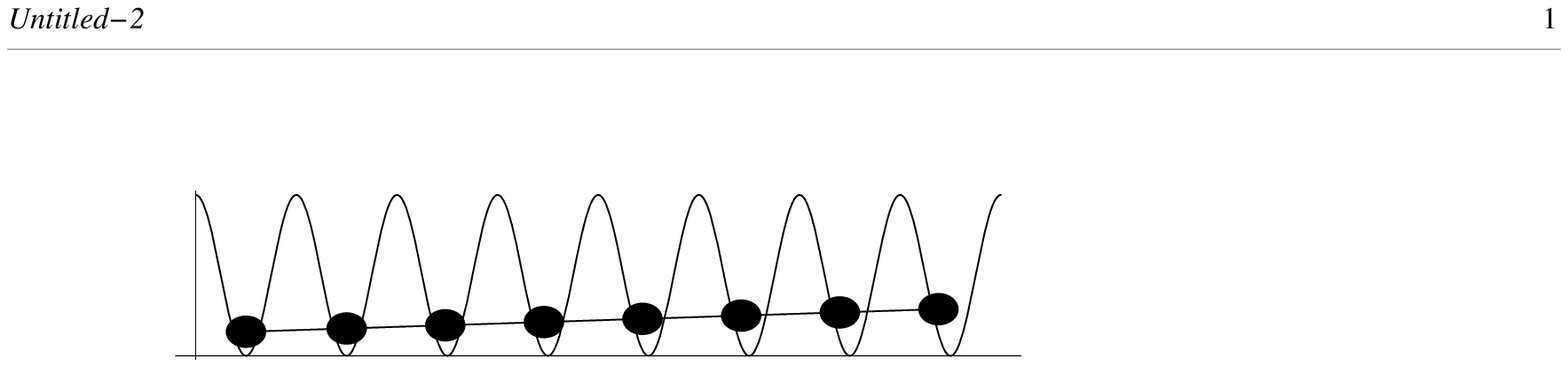,width=6.0cm,clip=TRUE,viewport=100 640 400 710,angle=0}}
\vspace{0.3cm}
\caption{Illustration of the effect of tensile misfit strain on structure.
The periodic potential has wavelength $a$, the lattice spacing set by the crystal,
while the natural length of the shorter doped Cu-Cu bonds is $l_0<a$.  Due to the tensile
misfit, the Cu atoms are unable to remain at the minima of the periodic
potential and, at the same time, to conform to the natural length $l_0$.}
\label{fig2}
\end{figure}

We assume that neighboring doped Cu sites in a stripe have a
separation, $l_0$ that is shorter than, but approximately equal to,
the average separation, $a$, dictated by the crystal structure and the
average doping level.  In this case, as the stripes form and increase
in length strain energy will build up.  The basic competition is
between the length scale of the periodic potential $a$ due to the
lattice structure and the natural length of the short bonds due to
near neighbor doped sites, as illustrated in Fig.~\ref{fig2}.
%
%
  The breakup
length scale of the stripes is approximated by comparing the strain
energy gain ($\sim k_{nn} N (a-l_0)^2/2$) which occurs when a
strained stripe of length $L= N a $ breaks up, to the formation energy
per bond of an unstrained stripe ($ - J_{nn}$).  This yields (dropping
constant prefactors)
\begin{equation}
N_c \approx ({J_{nn}\over k_{nn}}) ({1 \over a - l_0 })^2. 
\end{equation}
 The key parameters setting the scale of the energies in Eq. (1) are
 $k_{nn}$, which is proportional to the stiffness of the Cu-O bonds
 and $J_{nn}$ the energy gain (per bond) in forming stripes.  The
 spring constant $k_{nn} \sim V_0/ a^2$, where $V_0$ is of the order
 of an ionic bond energy ($\sim 5~eV$).  $J_{nn}$ is of the order of the
 stripe formation temperature, e.g., the temperature where the
 pseudogap appears $ \sim 400 K \sim 40 meV $. As discussed above, the
 misfit strain for cuprates is of order $(a-l_0)/a \sim 0.01$.
 Using these numbers in Eq. (1) indicates that for cuprates, the
 characteristic length of stripe domains, $N_c$, should be of order $100$
 lattice spacings.  The stripe length diverges quadratically (see
 Eq. (1)) as the misfit strain approaches zero leading to long range
 stripes in oxides with no misfit strain, as appears to be the case, for example,
in La\sb{2-x-y}Nd\sb{y}Sr\sb{x}CuO\sb{4}. 

We now develop a model to test whether the simple prediction (1) holds
when there are many stripes and to 
study the stripe nanostructures produced once
strain destabilizes long-range stripes.  This model is as simple
as possible while including the key aspects of doping, stripes and
misfit strain.   Note that this model considers only {\it stripe breakup} 
 and not stripe formation. We assume magnetic effects 
and long-range Coulomb interactions play a key role in
 stripe formation~\cite{stojk;prb00,seibo;prb98,bussm;pmb00}, but that
misfit strain leads to their breakup.  
Quantum fluctuations may also contribute to stripe breakup, 
however we consider here only misfit strain which we assume is
the dominant factor.
 The model is defined on a square lattice where each site is
assigned an occupancy number $n_i$ where $n_i=1$ indicates that the
site is doped with a hole and $n_i=0$ indicates that it is undoped.
\begin{eqnarray}
H = &-\sum_{ij}^{nn} J_{nn} n_i n_j + \sum^{nnn}_{ij}
 J_{nnn} n_i n_j + \mu \sum_i n_i\nonumber\\
&+ \sum_{ij}^{nn} k_{nn} (l_{ij} - l_0)^2 n_i n_j + \sum_i n_i V(\vec{r}_i).
\label{eq;ham}\end{eqnarray}
In the spirit of the stripes seen in
La\sb{1.6-x}Nd\sb{0.4}Sr\sb{x}CuO\sb{4} where a hole resides on every
second site along the stripe, we can simplify the computation by
considering a sublattice of every second Cu site
 and include an attractive nearest neighbor
and repulsive next nearest neighbor interaction, $J_{nn}$ and
$J_{nnn}$ respectively. These will result in stripe formation provided
$J_{nn}>0, J_{nnn}>0$.  The third term in the Hamiltonian has the
chemical potential, $\mu$, and allows us to vary the doping in the
model to some occupation density, $\rho$. Note that our sublattice
model only allows doping at every second copper site along the $x$ and
$y$ directions so that $\rho$ is related to the doping fraction $p$ in
the cuprates via, $\langle \rho \rangle \approx 4 p$. 
 The last two terms in Eq.~(\ref{eq;ham})
are the strain terms.  The first term contains the misfit strain which
occurs when two nearest neighbor sites on the sublattice are occupied
(this corresponds to next-nearest neighbor sites in the cuprates). The
parameter $l_0$ is the {\it natural length} of this bond, i.e., the
length that bond would take in the absence of stress.  The second term
$V(\vec{r}_i)$ is the periodic potential imposed on the copper oxide
layers by the crystal structure.  In our model we take this potential
to be modulated along both of the in-plane axes of the CuO\sb{2} planes
with a wavelength $a$, and to have a amplitude $V_0$.   These terms in
the Hamiltonian result in a finite stress energy unless $l_0=a$ as can
be seen in the schematic in Fig.~\ref{fig2}.  The strain
energy is long range and has to be relaxed self consistently using a
gradient or Newton's method.

\begin{figure}
\centerline{\epsfig{file=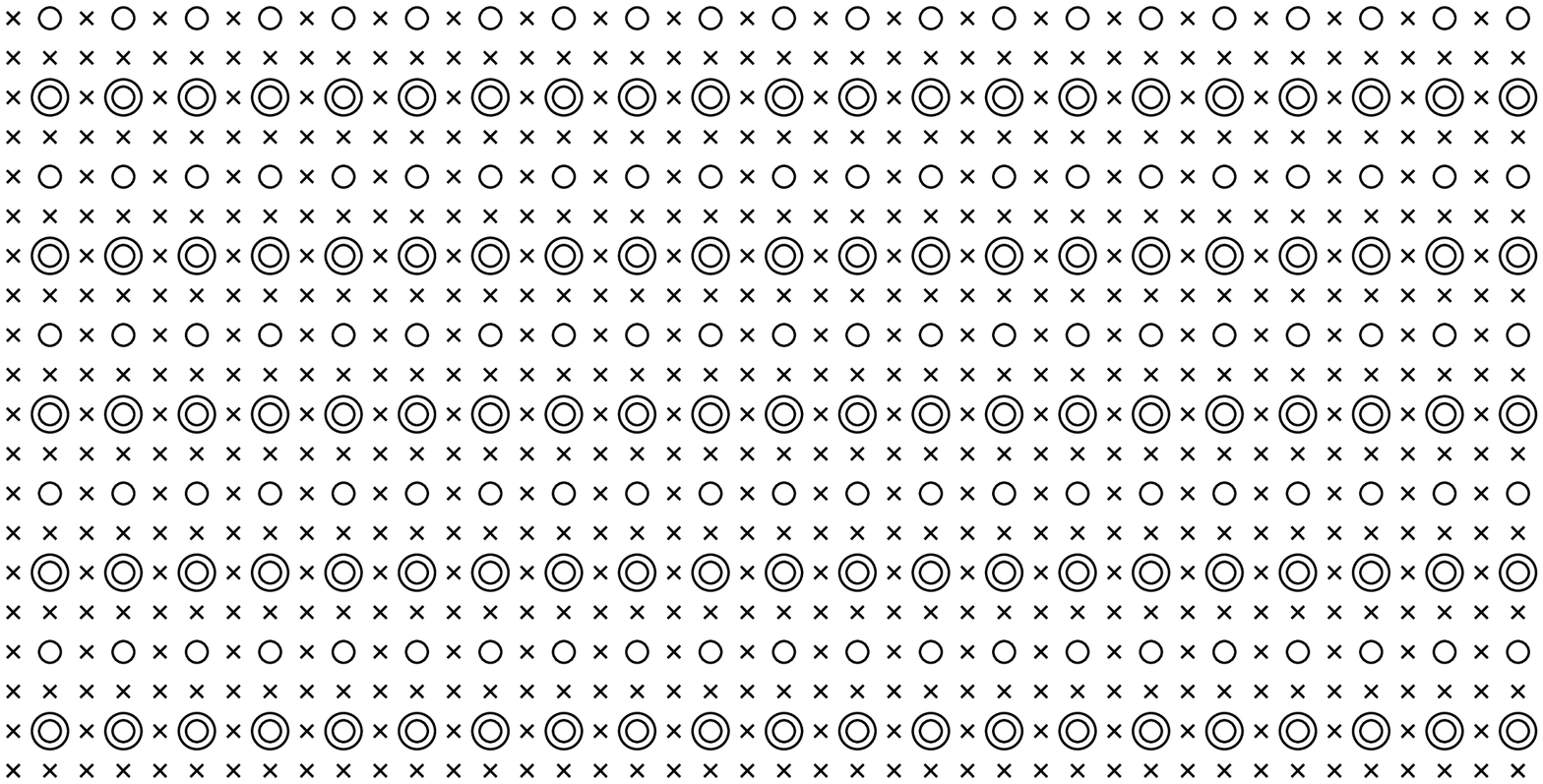,width=3.2cm,clip=TRUE,viewport=200 80 550 750,angle=270}}
\centerline{\epsfig{file=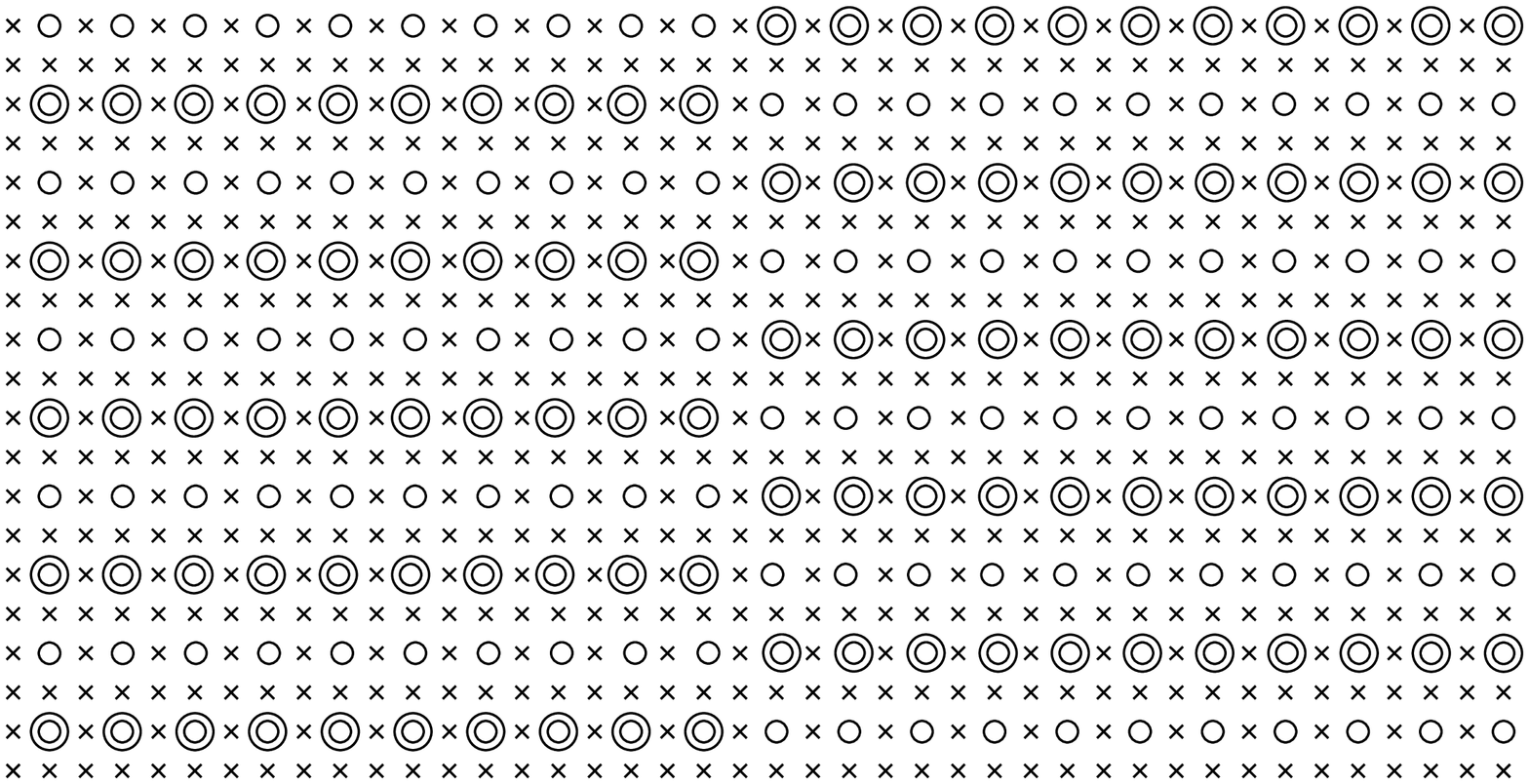,width=3.2cm,clip=TRUE,viewport=200 80 550 750,angle=270}}
\centerline{\epsfig{file=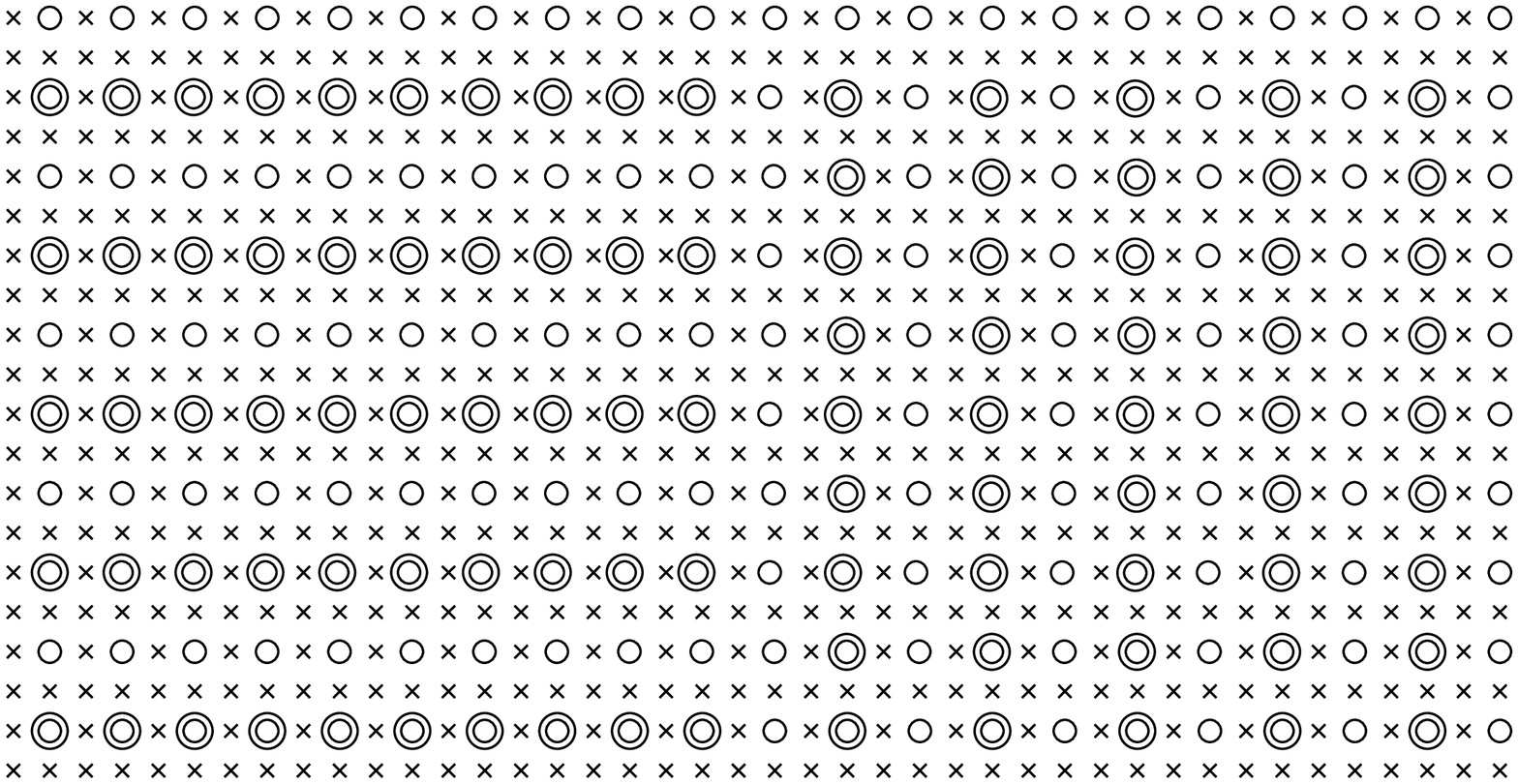,width=3.2cm,clip=TRUE,viewport=200 80 550 750,angle=270}}
\vspace{0.15cm}
\caption{Stripe nanostructures in cuprates.  In systems with 
tensile misfit, the static
stripe (top panel) has a high strain energy. Interleaved
stripes (middle panel) and the weave microstructure (bottom panel)
reduce this strain energy as they can accommodate strain relaxation.  One unit
cell of each microstructure is shown. Only copper atoms are shown. Concentric circles indicate doped
sites, while sites indicated by crosses are excluded from the strain relaxation.
 For illustration purposes the natural length of bonds  between doped sites are 10\% shorter 
than bonds between undoped sites. }
\label{fig3}
\end{figure}

It is difficult to find the ground state of 
the strained lattice gas (2), however it is easy to demonstrate
that long range stripes are unstable in the presence of misfit
strain and to test the general form (1).
To illustrate this, consider the three 
stripe nanostructures of Fig.~\ref{fig3}.
%
%
 From the lattice gas model (2), it is evident that the long range
 stripe has the lowest energy, provided there is no misfit strain.
 However in the presence of misfit strain, the lower two
 nanostructures of Fig.~\ref{fig3} have lower strain energy, though
 they do incur an energy cost proportional to $J_{nn}$ due to breaking
 up the long range stripes.  The strain energy in the long stripes is
 simply $k_{nn} (l_0 - a)^2/2$ per bond.  For a strain of $0.01$ and
 $V_0 = 1 = k_{nn}$(in lattice units), this leads to a strain energy of
 $0.0152$ for the structure of the top panel in Fig.~\ref{fig3}.  In
 contrast the interleaved stripe and weave structures (bottom two
 panels in Fig.~\ref{fig3}) have strain energies of $0.0100$ and $0.00765$
 respectively. These calculations were carried out using simple
 gradient descent on the unit cells of Fig.~\ref{fig3}, with the
 periodic cell shape fixed, and the boundary atoms fixed and excluded
 from the total energy calculation.  Tests using free and periodic
 boundaries and other nanostructures indicate that the state of lowest
 energy does depend on a variety of factors, including doping,
 boundary conditions and cell size.  However in all cases, the strain
 energy of the long range stripes is of order twice that of the stripe
 domain states and this is not sensitive to the modeling details.  It
 is thus reasonable to find the breakup length scale of the long range
 stripe array (top panel of Fig.~\ref{fig3}) by comparing the strain energy of
 long-range stripes with the nearest neighbor energy cost incurred in
 breaking the stripes.  This leads to $N_c k_{nn} (l_0 - a)^2 \approx
 J_{nn}$, which is consistent with the simple estimate (1). In
 general, the ground state structures of discrete strained systems are
 extremely
 rich~\cite{bak;rpp82,bratk;ptrsla96,alerh;prl88,march;jl33,frenk;zetf38}
 and we expect a similar richness in the ground structure of the
 Hamiltonian (2) as a function of doping and temperature, as will be
 described elsewhere.  However the main point emphasized here is that
 misfit strain destabilizes long range stripes in cuprates at a length
 scale well approximated by (1).

Now we briefly comment on a number of experimental observations that
can be explained by strain induced stripe nanostructures.  The most
important point is that the presence of strain will prevent the stripes
from ordering over long range resulting in broken stripes even at 
moderately high dopings.  Broken stripes will have significant
quantum fluctuations.   More structural
compliance in the form of enlarged octahedral tilt distortions will
reduce the misfit strain and cause the
stripes to grow in length, presumably slowing their dynamics.  Systems
where the CuO\sb{2} plane is in tension, or nearly in tension, such as
the Hg and Tl systems will have faster stripe
dynamics and therefore better superconducting properties and higher
\tc 's as observed.  Similarly, the increase in \tc\ with decreasing
tilt distortions~\cite{buchn;prl94,dabro;prl96} have an explanation
here.   The strained stripes will also tend to
stabilize the LTT phase over the LTO phase because this results in
greater structural compliance, as we discussed, and therefore a
decrease in the misfit strain.  Stripe microstructures such as those
shown in the third panel of Fig.~\ref{fig3} will result in {\it local}
LTT symmetry tilts but averaging over the two
90$^\circ$ rotated LTT variants will lead to effective LTO behavior, as observed in
La\sb{2-x}Ba\sb{x}CuO\sb{4}~\cite{billi;prl94}.  Also, severely
misfitting ions will give rise to large amplitude tilts locally,
resulting in locally varying structural compliance, which will tend to
pin the stripes.  This explains the observation that superconductivity
can be suppressed by increasing the mean-square dopant-ion size at
fixed average dopant-ion size~\cite{attfi;n98}.  Finally, we note that
ferroelastic effects~\cite{jung;unpub01} and the observation of tweed
microstructures~\cite{sheno;prb99,bratk;ptrsla96,schma;pml89}, as well as the
sensitivity of \tc\ on dopant ordering in
YBa\sb{2}Cu\sb{3}O\sb{6+\delta} and La\sb{2-x}CuO\sb{4+\delta}, might
have an explanation in strained stripe induced microstructures.

In summary, we point out that doping shortens the Cu-O bond length and that
inhomogeneous doping in the form of stripes results in a
tensile misfit strain. This breaks up the stripes and
results in, presumably fluctuating, domains of broken stripes where the
domain size depends on the misfit strain.  Structural compliance in the
form of buckling modes can reduce the misfit strain and result in
long range ordered and static or quasi-static stripes reducing or
destroying superconductivity.

SJLB thanks the National Science Foundation for support through grant
DMR-0075149 and appreciates stimulating conversations with Emil Bozin,
Takeshi Egami, Antonio Bianconi and Dragan Mihailovic.  PMD thanks the
Department of Energy for support through grant DE-FG02-90ER45418.


\begin{thebibliography}{10}

\bibitem{tranq;n95}
J.~M. Tranquada et al., 
\newblock Nature {\bf 375}, 561 (1995).

\bibitem{emery;pnas99}
V.~J. Emery et al., 
\newblock Proc. Natl. Acad. Sci. USA {\bf 96}, 8814 (1999).

\bibitem{marti;epl01}
I. Martin et al.,
\newblock Europhys. Lett. {\bf 56}, 849 (2001)

\bibitem{zaane;prb89}
J.~Zaanen and O.~Gunnarson,
\newblock Phys. Rev. B {\bf 40}, 7391 (1989).

\bibitem{white;prb00}
S.~R. White and D.~J. Scalapino,
\newblock Phys. Rev. B {\bf 61}, 6320 (2000).

\bibitem{stojk;prb00}
B. P. Stojkovi\'{c} et al.,
\newblock Phys. Rev. B {\bf 62}, 4353 (2000).

\bibitem{uehar;n99}
M.~Uehara et al. ,
\newblock Nature {\bf 399}, 560 (1999).

\bibitem{tranq;prl94}
J.~M. Tranquada et al.,
\newblock Phys. Rev. Lett. {\bf 73}, 1003 (1994).

\bibitem{tranq;prb96}
J.~M. Tranquada et al.,
\newblock Phys. Rev. B {\bf 54}, 7489 (1996).

\bibitem{kastn;rmp98}
M.~A. Kastner et al.,
\newblock Rev. Mod. Phys. {\bf 70}, 897 (1998).

\bibitem{mook;n99}
H.~A. Mook and {F. Do\v gan},
\newblock Nature {\bf 401}, 145 (1999);
H.~A. Mook et al., 
\newblock Nature {\bf 404}, 729 (2000).

\bibitem{arai;prl99}
M.~Arai et al., 
\newblock Phys. Rev. Lett. {\bf 83}, 608 (1999).

\bibitem{timus;rpp99}
T.~Timusk and B.~Statt,
\newblock Rep. Prog. Phys. {\bf 62}, 61 (1999).

\bibitem{mcque;prl99}
R.~J. McQueeney et al.,
\newblock Phys. Rev. Lett. {\bf 82}, 628 (1999).

\bibitem{lanza;n01}
A.~Lanzara et al.,
\newblock Nature {\bf 412}, 510 (2001).

\bibitem{goode;f92}
J.~B. Goodenough,
\newblock Ferroelectrics {\bf 130}, 77 (1992).

\bibitem{radae;prb94i}
P.~G. Radaelli et al.,
\newblock Phys. Rev. B {\bf 49}, 4163 (1994).

\bibitem{bozin;prl00}
E.~S. Bo{\v z}in et al., 
\newblock Phys. Rev. Lett. {\bf 84}, 5856 (2000);
ibid \newblock Phys. Rev. B {\bf 59}, 4445 (1999).

\bibitem{sheno;prb99}
S. R. Shenoy et al., 
\newblock Phys. Rev. B {\bf 60}, R12537 (1999);
K. O. Rasmussen et al. 
\newblock Phys. Rev. Lett. {\bf 87}, 055704 (2001).

\bibitem{thorp;jncs83}
M.~F. Thorpe,
\newblock J. Non-Crystalline Solids {\bf 57}, 355 (1983).

\bibitem{dove;b;lsfd98}
M.~T. Dove et al.,
\newblock in {\em Local Structure from Diffraction}, edited by S.~J.~L.
  Billinge and M.~F. Thorpe, page 253, New York, 1998, Plenum.

\bibitem{egami;js00}
T.~Egami and D.~Louca,
\newblock J. Supercond. {\bf 13}, 247 (2000).

\bibitem{axe;prl89}
J.~D. Axe et al.,
\newblock Phys. Rev. Lett. {\bf 62}, 2751 (1989).

\bibitem{white;prl98}
S.~R. White and D.~J. Scalapino,
\newblock Phys. Rev. Lett. {\bf 80}, 1272 (1998).

\bibitem{castr;prb01}
A. H. Castro-Neto \newblock Phys. Rev. B {\bf 64}, 104509 (2001).

\bibitem{caste;zpb97}
C Castellani et al., \newblock Z. Phys. B {\bf 103}, 137 (1997).

\bibitem{bussm;jpcm01}
A. Bussmann-Holder et al., \newblock J. Phys. Condens. Matter {\bf 13},
L169 (2001).


\bibitem{bianc;jpcm00}
A.~Bianconi et al.,
\newblock J. Phys. Condens. Matter {\bf 12}, 10655 (2000).

\bibitem{phill;pmb99}
J. C. Phillips 
\newblock Phil. Mag. B {\bf 79}, 527 (1999).


\bibitem{emery;prl00}
V.~J. Emery et al.,
\newblock Phys. Rev. Lett. {\bf 85}, 2160 (2000).

\bibitem{zaane;cm01}
J.~Zaanen et al.,
\newblock cond-mat/0102103  (2001).

\bibitem{seibo;prb98}
G. Seibold et al.,
\newblock Phys. Rev. B {\bf 58}, 13506 (1998).

\bibitem{bussm;pmb00}
A. Bussmann-Holder, et al., \newblock Phil. Mag. B {\bf 80}, 1955 (2000).

\bibitem{bak;rpp82}
P.Bak,
\newblock Rep. Prog. Phys. {\bf 45}, 587 (1982).

\bibitem{bratk;ptrsla96}
A.~M. Bratkovsky et al.,
\newblock Philos. Trans. Royal Soc. London A {\bf 354}, 2875 (1996).

\bibitem{alerh;prl88}
O.~L. Alerhand et al.,
\newblock Phys. Rev. Lett. {\bf 61}, 1973 (1988).

\bibitem{march;jl33}
V.~I. Marchenko,
\newblock JETP Lett. {\bf 33}, 381 (1981).

\bibitem{frenk;zetf38}
Y.~Frenkel and T.~Kontovova,
\newblock Zh. Eksp. Teor. Fiz. {\bf 8}, 1340 (1938).

\bibitem{buchn;prl94}
B.~B\"uchner et al.,
\newblock Phys. Rev. Lett. {\bf 73}, 1841 (1994).

\bibitem{dabro;prl96}
B.~Dabrowski et al.,
\newblock Phys. Rev. Lett. {\bf 76}, 1348 (1996).

\bibitem{billi;prl94}
S.~J.~L. Billinge et al.,
\newblock Phys. Rev. Lett. {\bf 72}, 2282 (1994).

\bibitem{attfi;n98}
J.~P. Attfield et al.,
\newblock Nature {\bf 394}, 157 (1998).

\bibitem{jung;unpub01}
J.~Jung,
\newblock preprint.

\bibitem{schma;pml89}
W.~W. Schmahl et al.,
\newblock Phil. Mag. Lett. {\bf 60}, 241 (1989).

\end{thebibliography}
%


\end{document}